\documentclass[aps,prb,showpacs,twocolumn]{revtex4}

\usepackage{graphicx}

\begin{document}

\title{Oxygen-covered tungsten crystal shape: time effects, equilibrium, surface energy and the edge-rounding temperature}

\author{Andrzej Szczepkowicz}

\affiliation{Institute of Experimental Physics, University of Wroclaw,
Plac Maksa Borna 9, 50-204 Wroclaw, Poland}

\date{\today}

\begin{abstract}
The equilibrium crystal shape (ECS) of oxygen-covered tungsten micricrystal is studied as a function of temperature.
The specially designed ultrafast crystal quenching setup with the cooling rate of 6000~K/s allows to draw conclusions about ECS
at high temperatures. The edge-rounding transition is shown to occur between 1300~K and 1430~K.
The ratio of surface free energies $\gamma(111)/\gamma(211)$ is determined as a function of temperature.
\end{abstract}

\pacs%
{%
 68.35.Md 
 65.40.gp	
 68.60.Dv 
 68.37.Vj 
}


\maketitle

\section{Introduction\label{sect-introduction}}

The study of equilibrium crystal shape (ECS) yields information on temperature-dependent surface free energy,
which is one of the fundamental quantities in surface physics and chemistry \cite{Bonzel2003}.
Understanding ECS is an important step in the ongoing challenge to construct
realistic models of solids. The problem of ECS for pure one-component crystals is
already complex (see eg. reviews \cite{Wortis1988,WilliamsBartelt1989,Bonzel2003}). Even more challenging
is the consideration of ECS with adsorption, which is important for the understanding of 
fundamental processes on catalyst surfaces \cite{FlytzaniStephanopoulosSchmidt1979}. 
There exist only few models and experimental reports concerning ECS with adsorption 
\cite{Shi1987, Shi1988, WilliamsBartelt1989, Hong2003, ChatainWynblattRohrer2005, SzczepkowiczBryl2005, NiewieczerzalOleksy2006}.

\begin{figure}
\includegraphics{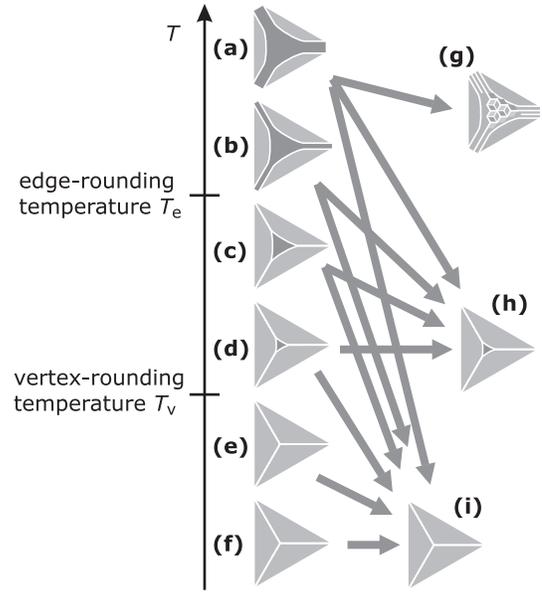}%
\caption%
{%
 \label{fig-evolution-general-scheme}%
 (a)--(f) The equilibrium crystal shape in the vicinity of the crystal vertex depends on temperature.
 (g)--(i) The actual crystal shape observed after quenching. The observed shape depends on the cooling
 rate and may differ from the equilibrium shape. During quenching small steplike facets may be formed as shown
 in (g).
}
\end{figure}

The general property of ECS is that it evolves with temperature. At absolute zero 
polyhedral shapes are predicted \cite{Wortis1988,FrenkenStoltze1999}, with atomically sharp vortices and atomically sharp edges. 
As the temperature is increased, entropic effects cause rough regions to appear on the crystal surface.
At a certain temperature the vortices and the edges become rounded. We will denote by $T_v$ the \textit{vertex rounding temperature}, and by $T_e$ the \textit{edge rounding temperature}. The dependence of ECS on temperature 
is illustrated in Fig.~\ref{fig-evolution-general-scheme}~(a)--(f).

Given the difficulty of experiments on ECS with adsorption, it is reasonable to choose an exprimentally simple
and well studied adsorption system -- O/W (for a review on O/W, see \cite{BrundleBroughton1990}). 
We treat O/W as a model system, which gives insight into general properties of thermal 
evolution of ECS predicted by theory. On the other hand, O/W is one of the candidate adsorption systems
for the fabrication of atomically sharp tungsten electrodes applied as practical electron and ion point sources
\cite{FuChengNienTsong2001,Lucier2005,Rezeq2006,BrylSzczepkowicz2006,Kuo2006,FujitaShimoyama2008,Rahman2008,
Chang2009, Kuo2008, Hommelhoff2009}, and understanding of the temperature dependence of 
O/W ECS allows one to control the sharpness of the electrode.

The adsorption of oxygen on tungsten is qualitatively different for low and high oxygen exposures. Thermal desorption spectroscopy studies reveal that after low exposures ($<$ 2 langmuir) adsorbate is all desorbed 
as O atoms, while with higher exposures oxygen is removed as tungsten oxides\cite{KingMadeyYates1971}. 
In the present work we concentrate on the low exposure case: 1.4 langmuir, where no alloying occurs.
The exposure of 1.4 langmuir corresponds to a coverage of less or equal to $5\cdot10^{14}$ molecules/cm$^2$.

Oxygen is known to increase the anisotropy of the surface free energy of tungsten and thus to induce pronounced
thermal faceting of the surface (see \cite{Szczepkowiczetal2005} and references therein). The O/W faceting
is observed in a very wide temperature range: it is already observed at 800 K\cite{Taylor1964}, 
while annealing at 1800~K still leaves enough oxygen on the surface for faceting to occur.
Oxygen adsorbate is mobile above 800~K -- it forms a two-dimensional lattice gas\cite{WangLu1983,WangLu1985}.

Oxygen-induced faceting of tungsten observed in experiments often takes the form of 
steplike (hill-and-valley) faceting, where steps or
pyramids/pits are formed on the surface. These non-convex shapes are steady-state, non-equilibrium forms.
Only under special conditions it is possible to observe a convex, equilibrium crystal form 
\cite{SzczepkowiczBryl2004, SzczepkowiczBryl2005} (global faceting). It is known that the
vertex-rounding transition occurs at $T_v=970\pm70$~K\cite{SzczepkowiczBryl2005}.

Still, there remain unsolved some interesting questions about the ECS of O/W crystals.
\begin{itemize}

\item As the temperature is increased, the transition from non-equilibrium steplike crystal surface to a convex
equilibrium shape may be observed \cite{SzczepkowiczBryl2004}. But as the temperature is increased further, it appears
that a reverse transition is taking place -- a single crystal edge is again disassembled into a non-convex group of
parallel edges (steplike faceting)\cite{SzczepkowiczBryl2004}, apparently violating theoretical predictions for ECS. 

\item With the present state of microscopic techniques, if one needs at the same time high spatial resolution ($\sim$1 nm), 
high sample temperatures ($\sim$1500~K) and good control of surface impurities, the only way to observe the ECS is to
observe the crystal after quenching (rapidly cooling) the crystal to a lower temperature. The question is: what is the necessary cooling rate for the atomic configuration to be preserved? Is the cooling rate applied in previous work sufficient?
(400--800~K/s\cite{SzczepkowiczBryl2004, SzczepkowiczBryl2005}) 

\item Is it possible possible to extract from microscopic observations 
any quantitative information about the dependence of the surface free energy on temperature?
(the problems are: microscope image deformations and doubts whether the cooling rate is sufficient).

\item Is it possible to observe the edge-rounding transition? (Not to be confused with the vertex-rounding transition observed before\cite{SzczepkowiczBryl2005}.)

\end{itemize}
The main purpose of the present paper is to answer the above questions. In this work the crystal is mounted 
on a special support allowing for extremely high cooling rates: 6000~K/s. As will be shown, for high crystal temperatures
this does make a difference.

\section{Experimental\label{sect-experimental}}

\subsection{The sample crystal}

The tungsten crystal had the form of a 2.6-mm long needle, 
with the cone half-angle not exceeding 10$^{\circ}$.
The axis of the needle was oriented along the [111] direction. 
The apex of the needle was approximately 
hemispherical, with the average radius of curvature 
 $270$~nm. 
In this work we focused on the changes in topography near the apex of the needle -- in the vicinity of the (111) pole
-- where the crystal shape is a good approximation of the equilibrium crystal shape.
The surface of the crystal was observed using
Field Ion Microscopy (FIM) \cite{FIM-Mueller-Tsong,FIM-Oxford}.

The base pressure of the vacuum chamber was $3\times10^{-10}$ Torr.
The experiments were carried out at fast pace to minimize
residual gas contamination. The whole process of crystal formation, 
from crystal cleaning to crystal quenching, took about 5 minutes.

The sample was carefully cleaned \textit{in situ} 
immediately before every experiment.
Cleaning by field evaporation, which is applied with great success
in FIM studies of low temperature single-atom
diffusion (see eg. \cite{AntczakEhrlich2005}), is not appropriate
for faceting studies, because at high temperature impurities
from the sample shank contaminate the field-evaporated tip apex.
For this reason we cleaned the sample thermally -- separately for each experiment.
This consisted of annealing at sufficiently high temperature
(1950--2200~K), where impurity-induced faceting is 
no longer observed. Surface cleanliness was verified by
performing ``blank'' experiments\cite{SzczepkowiczCiszewski2002}: 
the crystal was annealed at 1100 K without adsorbate, 
and no faceting was observed.

After cleaning, the tungsten crystal was cooled to 80~K and exposed to oxygen ($1.4\pm0.3$~Langmuir).
This provided the starting point of all experiments described in this work.

\subsection{Sample mounting for ultrafast cooling\label{sect-sample-mounting}}

\begin{figure}
\includegraphics[scale=.85]{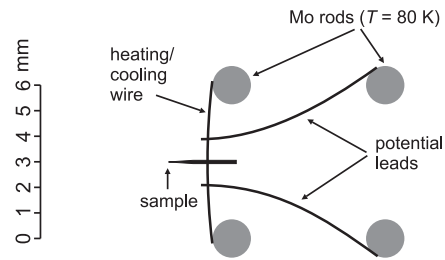}%
\caption%
{%
 \label{fig-sample}%
 Sample mounting for temperature control and ultrafast cooling.
}
\end{figure}

The main idea behind the ultrafast cooling setup constructed for this work is to heat the sample while maintaining
a very high temperature gradient in the sample support. When the heating stops, high temperature gradient causes
efficient conduction of heat from the sample, leading to a rapid decrease of the sample temperature. 
The~realization of this idea is shown in Fig.~\ref{fig-sample}. Four molybdenum rods ($\phi=1.5$~mm) are kept at low
temperature (80~K) and act as efficient heat sink. The sample is mounted on a short (6~mm) tungsten wire ($\phi=0.1$~mm),
which acts both as a resistive heater for the sample and as a heat conductor for sample cooling.
Additional two tungsten potential leads ($\phi=0.1$~mm) are used to measure the resistance, and thus to monitor the
temperature\cite{FIM-Mueller-Tsong}.

\begin{figure}
\includegraphics{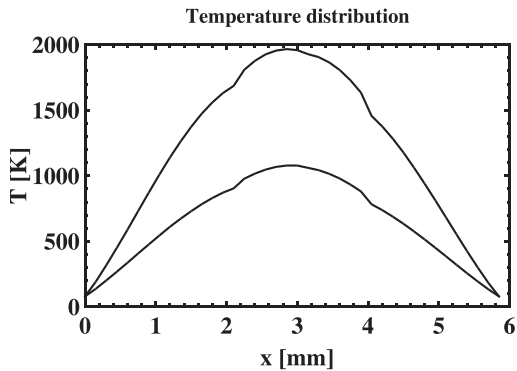}%
\caption%
{%
 \label{fig-temp-along-support-wire}%
 Calculated temperature distribution along the heating/cooling wire for sample temperatures 1070~K and 1880~K. 
 The potential leads act as additional heat sinks near $x=2$ and $x=4$~mm.
 Note the extremely high temperature gradient $dT/dx\sim500$~K/mm necessary for subsequent ultrafast cooling of the sample.
 }
\end{figure}

Figure \ref{fig-temp-along-support-wire} shows the resulting temperature distribution along the 6~mm long 
heating/cooling wire. Note that along the short 3~mm distance the temperature changes by almost 2000~K.
This high temperature gradient is desirable for ultrafast sample cooling, but it complicates
the measurement of the sample temperature. The standard method\cite{FIM-Mueller-Tsong} must be modified
to account for temperature gradients, as described in the next section.

\subsection{Determination of the temperature distribution and the cooling rate\label{sect-heat-equation}}

It is necessary to determine the spatial and temporal distribution of temperature
$T(x,t)$ in the sample support and the sample itself. Consider the segment $dx$
of a wire. 
The total rate of energy flow $dP$ into segment $dx$ is
equal to $c S d \frac{\partial T}{\partial t} dx$ ($c$ = specific heat,
$S$ = wire cross-section, $d$ = density, $\frac{\partial T}{\partial t}$ = rate of change of the temperature).
The total rate of energy flow $dP$ results from the balance of the supplied electric power
$dP_e$, thermally radiated power $dP_r$ and the power gain resulting from the gradient of 
the conducted power $dP_c=-\frac{\partial P_c}{\partial x}dx$:
$$
c S d \frac{\partial T}{\partial t} dx = dP_e - dP_r -\frac{\partial P_c}{\partial x}dx.
$$
The three contributions to $dP$ can be expressed as:
$dP_e=\frac{i^2\rho}{S}dx$ (Joule's law), 
$dP_r=\sigma a T^4 2 \sqrt{\pi S} dx$ (Stefan–-Boltzmann law), and
$\frac{\partial P_c}{\partial x}dx = \frac{\partial}{\partial x}( -S \lambda \frac{\partial T}{\partial x} ) dx$ (Fourier's law). We arrive at the following equation for the temperature $T(x,t)$:
$$
\frac{i^2\rho}{S} 
- \sigma a T^4 2 \sqrt{\pi S} 
+ \frac{\partial}{\partial x} \left( S \lambda \frac{\partial T}{\partial x} \right)
= c S d \frac{\partial T}{\partial t}.
$$
The material coefficients in this equation all depend on temperature:
electrical resistivity $\rho$, total emissivity $a$, thermal conductivity $\lambda$, sectific heat $c$, and density $d$.
We obtain this dependence from physical tables \cite{Ardenne1973,Hodgman1955,Espe,Washburn1930}.
On the basis of tabular data, we approximate the dependence of each parameter on temperature by a function of the form
$
f(T) = \sum_{i=-M}^{N} a_i T^i,
$
where $M,N$ are integers $\le5$. Only density is assumed constant over the entire temperature range.
Note also that the wire cross-section $S$ is a function of $x$ for the needle-shaped sample.

The wire configuration shown in Fig.~\ref{fig-sample} is quasi one-dimensional, so we use the above equation,
additionally assuming energy conservation and temperature continuity at the wire junctions. We solve the
equation for $T(x,t)$ by the finite-element method. An example stationary solution is shown in 
Fig.~\ref{fig-temp-along-support-wire}. To determine the crystal temperature, we measure the resistance of the
segment of the heating wire between the potential leads. On the other hand, this resistance is equal to
$\int \frac{\rho(T(x))}{S} dx$ and this value, on the basis of the full solution for $T$, can be correlated
with the temperature of the apex of the sample.

\begin{figure}
\includegraphics[width=86mm]{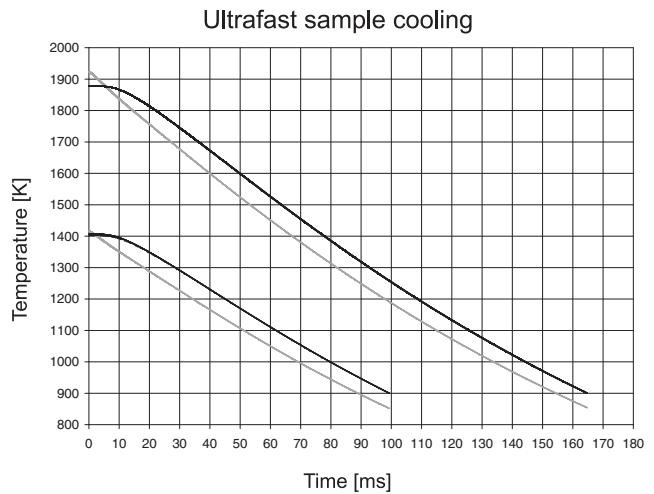}%
\caption%
{%
 \label{fig-wykresy-stygniecia}%
 The temperature of the sample apex (black) and of the sample welding point (gray) during sample cooling.
 }
\end{figure}

To determine the sample cooling rate, we calculate the non-stationary solution $T=T(x,t)$. An example 
result is shown in Fig.~\ref{fig-wykresy-stygniecia}. Here the heating current was turned off 
in a short time ($\sim$ 1~ms). To lower the cooling
rate, we program the power supply to reduce the current on a longer time scale.

\section{Results\label{sect-results}}

\subsection{Time effects -- the significance of the cooling rate\label{sect-results-time}}

In this section we demonstrate that surface topography observed after the crystal is quenched may depend on the cooling rate. The non-equilibrium structures formed during quenching should not be misteken for the equilibrium shape
(see Fig.~\ref{fig-evolution-general-scheme}). Both the general theory of equilibrium crystal shape 
\cite{Wortis1988, Bonzel2003}, as well as the solid-on-solid model of a crystal with adsorbate \cite{NiewieczerzalOleksy2006} predict that as the temperature of the crystal is increased, the vortices and the edges become rounded. Given the sufficiently high cooling rate, this rounding should be observed in the experiment. Indeed it has been shown that for the O/W crystal, the cooling rate in the range of 400--800 K/s is sufficient to observe the vertex rounding \cite{SzczepkowiczBryl2005}. However, this cooling rate was not sufficient to observe the edge rounding.

\begin{figure}
\includegraphics{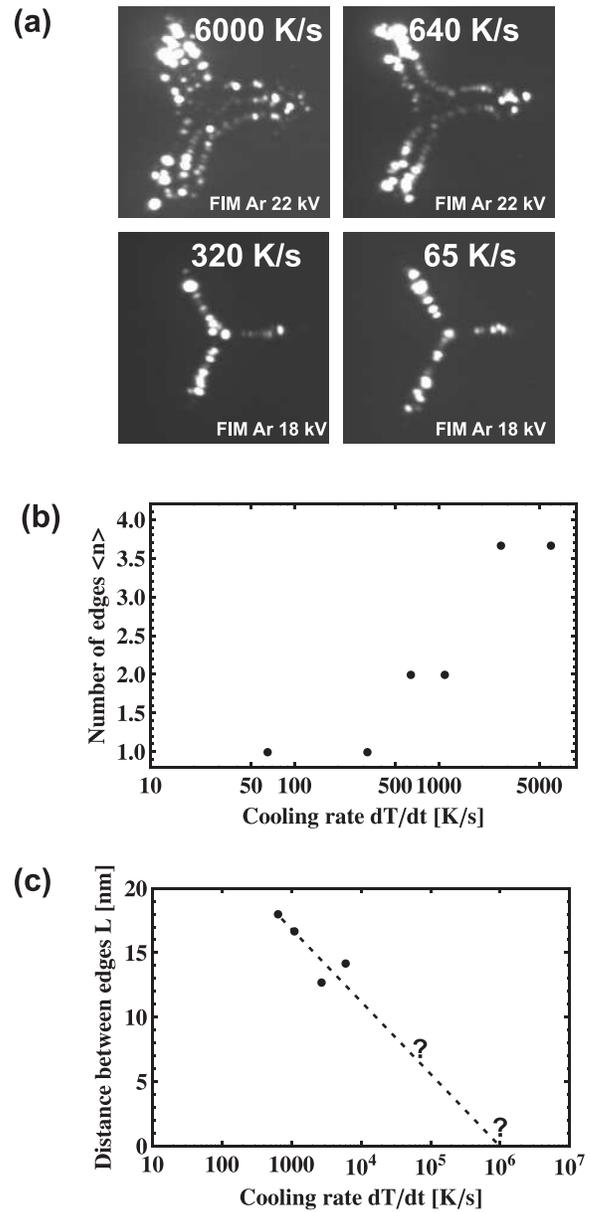}%
\caption%
{%
 \label{fig-time-effects}%
 Different steplike-faceted structures observed after annealing the crystal at a high temperature (1800~K) and cooling at different cooling rates.
 }
\end{figure}

The special sample setup described in Sect.~\ref{sect-experimental} allows us to explore the influence of the cooling rate on the observed crystal shapes. We start with a simple experiment of annealing the crystal at high temperature: 1800~K (similar results are also obtained for 1700~K). The shape observed after quenching depends on the cooling rate, as demonstrated in Fig.~\ref{fig-time-effects}. Ultrafast
cooling results in a steplike multiple-edge configuration corresponding to Fig.~\ref{fig-evolution-general-scheme}~(g).
Instead of a single crystal edge, we observe 3 groups of 3 or 4 parallel edges. The distance between the adjacent edges $L$
is smaller than 15~nm. The number of edges in a group $\langle n \rangle$ decreases to 2 for $dT/dt$ $\sim$ 1000~K/s.
For cooling rates lower than 500~K/s single crystal edges are observed (compare Fig.~\ref{fig-evolution-general-scheme}~(h)). Finally, for cooling rates lower than 100~K/s atomically sharp crystal 
vertex is observed (compare Fig.~\ref{fig-evolution-general-scheme}~(i)).

In Fig.~\ref{fig-time-effects} the strong dependence on the cooling rate is evident, and none of the observed shapes is the equilibrium shape. It turns out that cooling of the crystal as fast as 6000~K/s is still not sufficient for freezing
the high-temperature equilibrium shape. It is an interesting question what cooling rate would be sufficient. It is clear
that as the cooling rate increases, the number of steplike facets increases (Fig.~\ref{fig-time-effects}~(b)),
and the size of the facets decreases (Fig.~\ref{fig-time-effects}~(c)). The equilibrium shape would correspond
to $L=0$. It is difficult to extrapolate from only four experimental points, but it is probable that the
necessary cooling rate is from one to three orders of magnitude higher than the maximum rate available within the
present experimental setup.

\subsection{Equilibrium crystal shape\label{sect-results-ecs}}

As shown in Sect.~\ref{sect-results-time}, the atomic mobility of the substrate at 1800~K is so high 
that ECS is not preserved during quenching: small facets ($\sim$10~nm) are quickly formed ($\sim$100~ms),
so it is not possible at this temperature to observe ECS directly. However, for lower temperatures
surface diffusivity of the substrate is moderate and ECS can be observed. But if the temperature is too low,
the opposite problem emerges: material mobility is insufficient to overcame the kinetic barriers.
As demonstrated previously\cite{SzczepkowiczBryl2004, SzczepkowiczBryl2005}, it is best to proceed as follows:
first the crystal should be annealed at 1500~K, where hill-and-valley faceting no longer occurs and a globally faceted
shape emerges, and then the crystal should be annealed at the desired equilibration temperature (80~s is sufficient 
for a crystal of $r=270$~nm for $T > 1000$~K).

\begin{figure}
\includegraphics{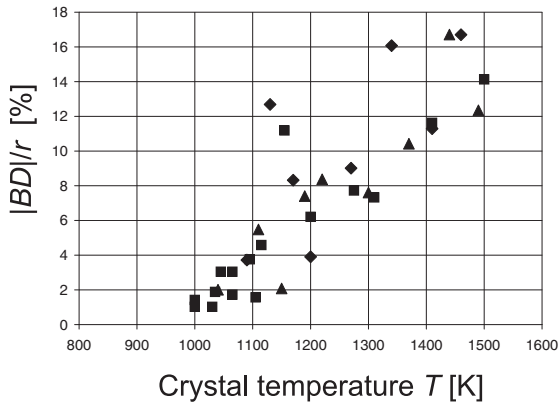}%
\caption%
{%
 \label{fig-T-d-plot}%
 The distance between the edge end points $|BD|$ (see Fig.~\ref{fig-gamma-calculation-drawing}), corrected for microscope image distortions \cite{NiewieczerzalOleksySzczepkowicz2010}, expressed as percent of crystal radius $r$. Squares denote previous data \cite{SzczepkowiczBryl2005} obtained for 
$dT/dt$ = 400--800~K/s. Diamonds and triangles denote current results for $dT/dt$ = 6000~K/s, with the equilibration 
annealing temperature approached from below (diamonds) and from above (triangles). Five points deviate from the main trend
problably due to contamination of the sample surface and will be discarded in energy calculations.
}
\end{figure}

What happens when the cooling rate 400--800~K/s applied previously\cite{SzczepkowiczBryl2005} is increased by an order of magnitude?  
The results shown in Fig.~\ref{fig-T-d-plot} demonstrate that this makes no difference in the temperature range
1000--1300~K. The results for $T>1300$~K are not conclusive: on the average, ultrafast cooling leads
to a slightly higher $|BD|$, as could be expected, but the effect is of the order of the experimental error. 
The main factors contributing to the fluctuations of the measured $|BD|$ are 
fluctuations of the residual contamination of the crystal surface, and fluctuations of the oxygen coverage.

An important property of ECS is that it does not depend on the sample history, but only on the
final annealing temperature (reversibility). To verify this, we performed half of the experiments by approaching the
equilibration temperature from lower temperatures (800~K), an the other half -- from higher temperatures (1500~K).
In both cases the starting point was the globally faceted crystal as described above. The result is shown
in Fig.~\ref{fig-T-d-plot}. Within the experimental error, the two procedures lead to the same crystal shape -- the ECS.

\begin{figure}
\includegraphics[width=86mm]{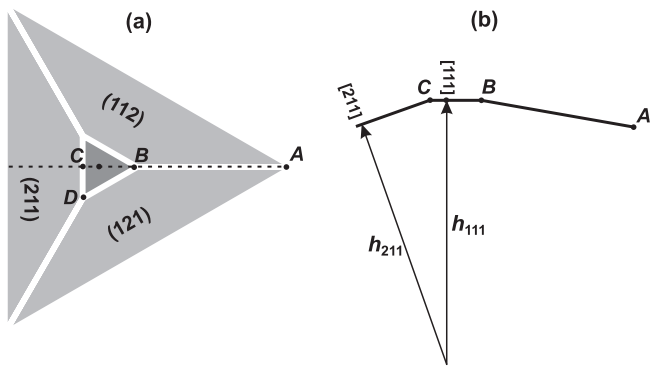}%
\caption%
{%
 \label{fig-gamma-calculation-drawing}%
 (a) Crystal surface in the vicinity of the [111] crystal pole forms a three-sided truncated pyramid.
 (b) Crystal cross-section assumed for surface free energy calcultations.
}
\end{figure}

\subsection{Calculation of the surface free energy}

As the temperature is increased, the (111) region increases at the expense of the (211) regions. According to the theory of the equilibrium crystal shape \cite{Herring1951, Wortis1988}, this means that the surface free energy of the (111) surface ($\gamma_{111}$) decreases relative to the surface free energy of the (211) surface ($\gamma_{211}$). The experimental data allow to calculate the ratio of surface free energies $\gamma_{111}/\gamma_{211}$ as a function of crystal temperature. In order to calculate this ratio, we assume that the crystal has a simple, perfect form depicted in Fig.~\ref{fig-gamma-calculation-drawing}. According to the Wulff theorem we have 
$\gamma_{111}/\gamma_{211}=h_{111}/h_{211}$. In the following paragraphs we describe how the ratio $h_{111}/h_{211}$ can be experimentally determined. 

First we determine the average crystal radius $r$ from the formula well known in field electron/ion microscopy \cite{Gomer1993, FIM-Oxford}: $E=V/kr$, where $E$ is the ``best image field'' (22 V/nm for argon imaging), $V$ the ``best image voltage'',
$r$ the crystal radius, and $k\approx5$ (for microscope samples annealed at temperatures exceeding 2000~K). 
The calculated crystal radius is 270~nm.

In the next step, we analyse the FIM image of the crystal annealed at temperature $T$ and determine $|BD|$, the distance between
the edge end points. This would be easy if the microscope maginfication was uniform -- but it is not \cite{FIM-Oxford}, and 
for faceted crystals local magnification may vary by as much as an order magnitude\cite{NiewieczerzalOleksySzczepkowicz2010}.
We use the results of very complex and time consuming ion trajectory calculations to calculate the local enhancement
of the microscope magnification near the crystal vertex\cite{NiewieczerzalOleksySzczepkowicz2010}. The local magnification factor turns out to be 1.14--3.26 for crystal shapes formed at temperatures 1000--1500~K. The calculated distances $|BD|$ 
fall in the range 2.8~nm--38~nm.

Once $r$ and $|BD|$ are known, the ratio $h_{111}/h_{211}$ can be determined from geometry -- see Fig.~\ref{fig-gamma-calculation-drawing}. Up to the vertex rounding temperature, the crystal forms a sharp three-sided
pyramid, and the distances $|BD|$ and $|BC|$ are zero. In this special case $h_{111}/h_{211}=1/\cos\alpha$, where
$\alpha$ is the angle between [111] and [211] directions: $\alpha=\arccos(\frac{2\sqrt{2}}{3})$. For higher crystal temperatures the pyramid is truncated and $|BC|=\frac{\sqrt{3}}{2}|BD|$. We assume that $h_{211}$ remains constant and equal to the average crystal radius $r$, while $h_{111}$ decreases from the maximum value of $r/\cos\alpha$ to a value of
$r/\cos\alpha - |BC|/(\cot\alpha+\cot\beta)$, where $\beta$ denotes the angle between the segments $AB$ and $BC$:
$\beta=\pi/2-\arccos(\frac{1}{\sqrt{33}})$. Consequently, we obtain the following formula for the ration of the surface free energies:
$$
\frac{\gamma_{111}}{\gamma_{211}}=
\frac{h_{111}}{h_{211}}=
\frac{r/\cos\alpha-|BC|/(\cot\alpha+\cot\beta)}{r},
$$
which simplifies to
$$
\frac{\gamma_{111}}{\gamma_{211}}=
\frac{3}{2\sqrt2}-\frac{\sqrt3}{12\sqrt2}\frac{|BD|}{r},
$$
where $|BD|$ is the distance between the edge end points corrected for the ion trajectory deflections,
and $r$ is the average crystal radius. 

\begin{figure}
\includegraphics{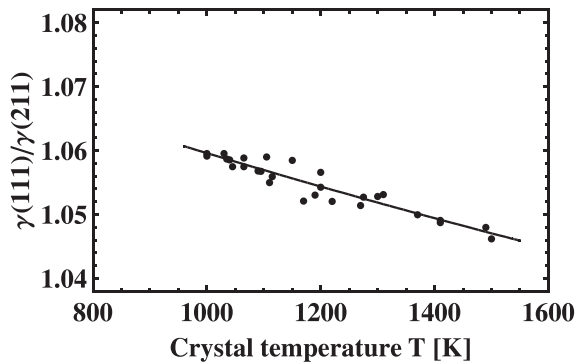}%
\caption%
{%
 \label{fig-surface-free-energy-plot}%
 The ratio of surface free energies $\gamma_{111}/\gamma_{211}$ as determined from experiment.
  }
\end{figure}

The resulting ratio of surface free energies $\gamma_{111}/\gamma_{211}$ as a function of crystal temperature is plotted in Fig.~\ref{fig-surface-free-energy-plot}. 
When the ratio is $1/\cos\alpha=\frac{3}{2\sqrt2}=1.061$, the crystal is at the vertex rounding temperature ($T_0=960$~K).
The ratio decreases to 1.052 at 1300~K. This means that the anisotropy of
the surface free energy decreases with increasing temperature, in accord with the theory of ECS \cite{Wortis1988,WilliamsBartelt1989}.

\subsection{The edge-rounding temperature\label{sect-results-er}}

\begin{figure}
\includegraphics{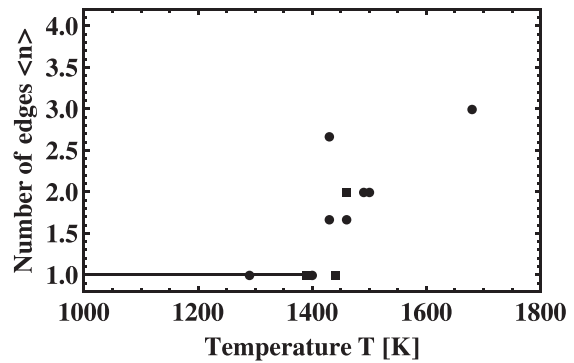}%
\caption%
{%
 \label{fig-edge-rounding-temperature}%
 A crystal with single sharp edges is annealed at temperature $T$ for time $t=10$~s (circles) or $t=80$~s (squares).
 Annealing above 1430~K destroys the crystal edge.
}
\end{figure}

If we extrapolate the results shown in Fig.~\ref{fig-time-effects} to infinite cooling rate,
we obtain an infinite number of infinitely small steps -- that is, an equilibrium configuration with a rounded edge.
This means that the crystal annealed at 1800~K (Sect.~\ref{sect-results-time}) is above the
edge-rounding transition: $1800~\mathrm{K} > T_e$. Even with the present ultrafast cooling setup,
it is not possible to observe the edge-rounding transition directly -- during quenching from high temperatures
the smooth edge reconstructs into steplike facets
(as predicted theoretically -- see in Figs.~7 and 8 of Ref.~\cite{NiewieczerzalOleksy2006}). 
However, with the present setup 
it is possible to estimate the edge rounding temperature $T_e$ in the following way.
First, the crystal is annealed at 1700~K (80~s), then the temperature is slowly lowered to 80~K (during $\sim$100~s)
to allow for the formation of single sharp edges. The crystal with sharp edges is then annealed at a temperature $T$,
to determine whether this temperature will destroy the sharp edge. Annealing below $T_e$ preserves the sharp edge.
Annealing slightly above $T_e$ results in edge rounding, but the single edge may be rebuilt during quenching.
Annealing at higher temperatures results in extensive edge rounding and steplike faceting of the rounded region during quenching. The results are shown in Fig.~\ref{fig-edge-rounding-temperature}. From these results we can estimate
the edge-rounding temperature from above: $T_e<1430$~K. The lower bound for $T_e$ is 1300~K on basis
of the results of Sect.~\ref{sect-results-ecs} (up to 1300~K freezing is fast enough). In conclusion, 
$1300~\mathrm{K}<T_e<1430~\mathrm{K}$.

\section{Discussion}

\subsection{The significance of the cooling rate}

The concept of surface free energy is most often called upon in two contexts: the equilibirum crystal shape
and the thermal faceting of surfaces. It is best when the experiments are performed by observing the crystal
at the equilibriation temperature. As mentioned in Sect.~\ref{sect-introduction}, for technical reasons this is not always
possible, and great care must be taken to ensure that no significant reconstruction occurs when the crystal is quenched for
observation. In Sect.~\ref{sect-results} we demonstrated that the possible effect of the finite cooling rate can be examined
by varying the cooling rate.

In many of the thermal faceting experiments reported in the literature, the observation of the surface is performed
after cooling of the crystals, and the cooling rate is not given. At the same time the experimental results are often
discussed in terms of zero-temperature models. By coincidence, this approach is often satisfactiory. The crystal shape formed during cooling at moderate cooling rates is close to the crystal form characteristic for the temperature where
massive surface diffusion stops (for metals near 25\% of the crystal melting temperature), and this may be close to the theoretical crystal form at absolute zero, with sharp edges and sharp vortices. However, it is a good experimental practice to pay attention to the cooling rate and to watch for its possible influence. For example, Song et al. \cite{Song1995} 
measured the cooling rate and showed by high-temperature LEED observation that 60~K/s available in their setup is insufficient for freezing the Pd/Mo crystal form at 900~K. The cooling rate in our setup is two orders of magnitude higher. As shown
in Sect.~\ref{sect-results}, this is sufficient for freezing the O/W crystal at moderate temperatures, but still not sufficient
for freezing the ECS at 1800~K.

The present cooling rate could be further increased by shortening the heating/cooling wire (see Fig.~\ref{fig-sample}).
Shortening of the wire not only reduces heat capacity, but also reduces thermal resistance of the wire, thus
leading to a strong increase of the quenching rate. This is also seen from the scaling properties of the equation for $T(x,t)$ derived in Sect.~\ref{sect-heat-equation}. During cooling, electric current is switched off (drop the first term of the equation), and radiation plays a smaller role than heat conduction (drop the second term). The equation is second order in $x$ and first order in $t$. If $T_1(x,y)$ is a solution, so is $T_2(x,y)=T_1(ax,a^2t)$. This means that if the wire is $a$ times shorter (making the temperature gradient $a$ times larger), the cooling proceeds $a^2$ times faster.
On the other hand, the influence of the wire cross section $S$ on the quenching rate is much weaker, because decreasing of the cross section area reduces heat capacity of the wire, but increases its thermal resistance, and the two effects cancel out
(if radiation is neglected, $S$ on both sides of the equation cancels out).

\subsection{Comparison with the theories of ECS}

The temperature dependence of ECS for oxygen-covered tungsten crystal follows the general pattern predicted for crystals
without adsorbate \cite{Wortis1988,WilliamsBartelt1989,Bonzel2003,FrenkenStoltze1999}. Increasing the temperature
makes the surface free energy and the ECS more isotropic. At low temperature sharp edges and vortices are observed, 
and at elevated temperature the vortices and the edges became rounded. For O/W crystal the vertex-rounding transition
occurs at $T_v=960$~K. In Sections~\ref{sect-results-time} and \ref{sect-results-er} we present evidence
that the edge-rounding transition also occurs, and estimate its temperature: $1300~\mathrm K<T_e<1430~\mathrm{K}$. The increasing isotropy
of the surface free energy is quantitatively shown in Fig.~\ref{fig-surface-free-energy-plot} -- the ratio of
the surface free energies $\gamma(111)/\gamma(211)$ decreases from 1.061 at 960~K to  1.052 at 1300~K.

There exist only few models and experimental reports concerning ECS with adsorption 
\cite{Shi1987, Shi1988, WilliamsBartelt1989, Hong2003, ChatainWynblattRohrer2005, SzczepkowiczBryl2005, NiewieczerzalOleksy2006}. The solid-on-solid model developed by Niewieczerzal and Oleksy \cite{NiewieczerzalOleksy2006}  predicts the temperature dependence of an adsorbate-covered crystal shape. Although the parameters of their model were adjusted for palladium-covered tungsten, the adsorption systems Pd/W and O/W exhibit certain similarities \cite{Szczepkowiczetal2005},
and the model shows qualitative agreemenent with the present work:
(a) To overcame the kinetic barriers and obtain ECS, the crystal must be first heated to a temperature where vortices
and edges became rounded (Niewieczerzal and Oleksy call this ,,the defaceting temperature'' \cite{NiewieczerzalOleksy2006}).
(b) At low temperature the ECS is polyhedral, at high temperature vortices and edges are smoothly rounded.
(c) Finite cooling rates at high temperatures lead to formation of non-equilibrium steplike facets.

\subsection{Implications for faceting on flat crystals}

The ECS is related to the problem of thermodynamic stability of flat crystal surfaces. 
Herring \cite{Herring1951} showed that a flat crystal surface is stable if and only if the corresponding crystal
orientation is present on the ECS. If a facet of orientation $(hkl)$ or a rounded vertex of orientation $(hkl)$ 
is present on the ECS, then a flat crystal of orientation $(hkl)$ is thermodynamically stable. On the other hand,
a sharp vertex implies that the corresponding flat surface is unstable and steplike (hill and valley) faceting will be observed.
As the ECS depends on temperature, the stability of some flat surfaces is also temperature-dependent.

Song and coworkers \cite{Song1995,Liaoetal2007} observed faceted/planar phase transitions on flat crystals for
Pd/Mo(111) (at $T=850$~K), O/Mo(111) (at $T=950$~K) and Pd/W(111) ($\sim$1130~K, complicated by partial desorption of Pd). 
According to Herring's theorem these correspond to vertex-rounding transitions for the crystal vertex in the [111] direction.
Unfortunately no experiments were carried out for the ECS of these adsorbate-covered crystals. Palladium adsorption 
and thermal faceting was studied on curved tungsten surfaces \cite{SzczepkowiczCiszewski2002}, but it turns out that
a crystal of radius 200~nm cannot be thermally equilibrated in the temperature range where the adsorbate remains on
the surface -- kinetic constraints prevent the crystal from reaching a convex, globally-faceted equilibrium shape.

It is interesting that the reversible planar-faceted transition is observed for O/Mo(111) at $T=950$~K \cite{Song1995}, 
which coincides with the vertex-rounding temperature $T_v=960$~K for O/W for the [111] vertex. This is unexpected,
because the melting point of Mo is $\sim$20\% lower then the melting point of W.
Perhaps it is not the substrate properties that govern the vertex-rounding process, but some transition 
in the two-dimensional lattice gas formed by the adsorbed oxygen. Studies of oxygen on different substrates could clarify this matter.

According to Herring's theorem, the present results have following implications for oxygen-induced faceting 
of flat tungsten crystals. At the vertex-rounding temperature $T_v=960$~K, a flat O/W(111) crystal is expected
to undergo a transition from \{211\}-faceted to defaceted phase, similarly to a transition observed by Song et al.
for O/Mo(111) \cite{Song1995}. A similar conclusion can be drawn on basis of the edge-rounding transition.
At the edge rounding temperature, the edge between (121) and (211) facets becomes rounded, and this means that
the orientation (332), hich is halfway between (121) and (211), becomes present on the ECS. For this reason
at the edge-rounding temperature $T_e$ ($1300~\mathrm K<T_e<1430~\mathrm{K}$) a flat O/W(332) crystal is expected
to undergo a transition from \{211\}-faceted to defaceted phase.

\section{Conclusions}


Using an ultrafast cooling setup, we studied the equilibrium crystal shape of oxygen-covered (1.4~Langmuir) tungsten microcrystal (average radius 270~nm) in the vicinity of the (111) crystal pole. The results can be summarised as follows:

\begin{itemize}

\item Up to 1300~K, cooling rates in the range $\sim$600--6000~K/s are sufficient for the observation of the ECS -- no observable changes occur in the crystal shape during cooling. On the other hand, we presented evidence that
the ECS at 1700--1800~K is not preserved during cooling even at 6000~K/s -- non-equilibrium steplike (hill-and-valley) facets
are formed.

\item The ECS undergoes vertex-rounding transition at $T_v=960$~K and edge-rounding transition at 
$1300~\mathrm K<T_e<1430~\mathrm{K}$.

\item The ratio of temperature-dependent surface free energies
$\gamma(111)/\gamma(211)$ decreases from 1.061 at 960~K to 1.052 at 1300~K.

\item A faceting/defaceting transitions are expected to occur for flat O/W crystals:
at 960~K for (111) orientation, and between 1300~K and 1430~K for (332) orientation.

\end{itemize}

\begin{acknowledgments}
I am grateful to Dr. Czeslaw Oleksy for helpful discussions.
\end{acknowledgments}



\end{document}